\documentclass[prb,twocolumn,aps,showpacs,floatfix]{revtex4}

\usepackage{graphicx}
\usepackage{bm}
\usepackage{amssymb}
\usepackage{amsmath}
\usepackage{subfigure}
\usepackage{tikz}
\usetikzlibrary{shapes,arrows}
\usepackage{amsmath}

\newcommand{\tab}{\hspace*{2em}}
\newcommand\T{\rule{0pt}{2.6ex}}
\newcommand\B{\rule[-1.2ex]{0pt}{0pt}}

\begin{document}

\title{Strong coupling expansion for bosons on the kagome lattice}
\author{Vipin Kerala Varma and Hartmut Monien}
\affiliation{Bethe Center for Theoretical Physics, Universit\"{a}t Bonn, Germany\\}
\date{\today}
\begin{abstract}
We use series expansion techniques for analyzing properties of the phase transition between the Mott insulating
and superfluid phase for bosons on the kagome lattice, and the multicritical point in the ground-state phase diagram for unit-filling is calculated. It is seen that of the clusters that contribute with non-zero weights to the ground state energy, many contain rings. The exponential decay coefficients of ground state correlations are also obtained within the Mott phase. For excited properties, quasiparticle dispersion and effective masses for particles and holes are computed. Furthermore at $8^{\textrm{th}}$ order, the coherence-length critical exponent $\nu$ is found to be comparably close to that of the 3D XY model.

\end{abstract}

\maketitle

\begin{figure}[bbp]
\centering
\subfigure[]{\label{kaglatticereal}
\begin{tikzpicture}[scale=0.75]
 \draw[very thick] (0,0) -- (1,0);
 \draw[->, very thick] (1,0) -- (2,0);
 \draw (2,0) -- (3,0);
 \draw (1,0) -- (1.5,-0.866);
 \draw (1.5,-0.866) -- (2,0);
\draw [very thick] (0,0) -- (0.5,0.866); 
 \draw (0.5,0.866) -- (1,0);
 \draw (2,0) -- (2.5,0.866);
 \draw (2.5,0.866) -- (3,0);
 \draw (0.5,0.866) -- (0,1.732);
 \draw [->, very thick] (0.5,0.866) -- (1,1.732);
\draw (0,1.732) -- (1,1.732); 
 \draw[dashed] (1,1.732) -- (2,1.732);
 \draw[dashed] (1,1.732) -- (1.5,2.598);
 \draw[dashed] (1.5,2.598) -- (2,1.732);
 \draw (2,1.732) -- (3,1.732);
 \draw (2,1.732) -- (2.5,0.866);
\draw (3,1.732) -- (2.5,0.866); 
\draw (0,0) circle (1mm);
\draw (1,0) circle (1mm);
\draw (2,0) circle (1mm);
\draw (3,0) circle (1mm);
\draw (1.5,-0.866) circle (1mm);
\draw (0.5,0.866) circle (1mm);
\draw (2.5,0.866) circle (1mm);
\draw (0,1.732) circle (1mm);
\draw (1,1.732) circle (1mm);
\draw (2,1.732) circle (1mm);
\draw (3,1.732) circle (1mm);
\draw (1.5,2.598) circle (1mm);
\end{tikzpicture}
}\tab
\subfigure[]{\label{kaglatticemom}
\begin{tikzpicture}[scale=0.75]
\draw (-1.0471,1.8137) -- (0, 1.8137);
\draw (0, 1.8137) -- (1.0471,1.8137) node[above right]{K ($\frac{\pi}{3}, \frac{\pi}{\sqrt{3}}$)};
\draw (1.0471, 1.8137) -- (1.5707,0.9068) node[right]{M};
\draw (1.5707,0.9068) -- (2.0943,0) node[right]{X ($\frac{2\pi}{3}, 0$)};
\draw (2.0943,0) -- (1.0471,-1.8137);
\draw (-1.0471,-1.8137) -- (1.0471,-1.8137);
\draw (-1.0471,-1.8137) -- (-2.0943,0);
\draw (-2.0943,0) -- (-1.0471,1.8137);
\draw [dashed] (0,0) node[left]{$\Gamma$} -- (1.0471, 1.8137);
\draw [dashed] (0,0) -- (2.0943,0);
\draw (0,1.8137) circle (0.2mm);
\draw (1.0471,1.8137) circle (0.2mm);
\draw (1.5707,0.9068) circle (0.2mm);
\draw (2.0943,0) circle (0.2mm);
\draw (0,0) circle (0.2mm);
\end{tikzpicture}
}
\caption{(a) Kagome lattice in real space with basis vectors (bold) and unit cell (dotted). (b) Reciprocal lattice with symmetry points labeled.}
\label{lattice}
\end{figure}
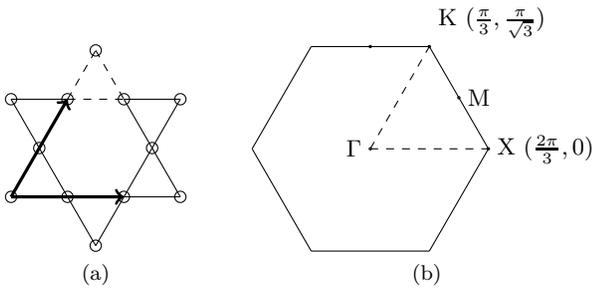

Strong coupling expansion has proved to be one among many powerful techniques for providing quantitative descriptions of strongly correlated bosonic systems. Even at low order\cite{Freericks}, the disctinction between 1D and higher dimensional critical systems was clearly revealed. This was confirmed in a DMRG study\cite{Kuehner} and a high order series expansion\cite{ElstnerMonien}. Prior to this, an infinite range mean-field analysis\cite{Fisher} had proposed the various possible phases in such systems and a later local mean-field study\cite{Krishna} provided a quantitative description of the phase diagram. Furthermore, the suggestion that ultracold dilute gases of bosonic atoms could be modelled by such systems\cite{Jaksch} and the subsequent experimental realization of the transition from the Mott insulator (MI) to the superfluid (SF) phase\cite{Greiner}, renewed interest in the study of the field within the context of cold atoms.

The current work analyzes such bosonic Hubbard models on the kagome lattice (see Fig. \ref{lattice}) using the series expansion technique\cite{SinghGelfand}, which has been applied on the 2D square lattice\cite{ElstnerMonien} and the 2D triangular lattice\cite{ElstnerMonienArxiv}. The present study attempts to quantify the bosonic model's characteristic features on the kagome lattice, complementing earlier studies of similar systems\cite{Isakov, Sengupta, Huber, Murthy, Damski, SinghHuse}. Despite the enormous bosonic Hilbert space, the similarity transformation\cite{Gelfand} can be used to treat such systems by creating effective Hamiltonians; in addition, there are no finite size effects since the thermodynamic limit is taken care of right at the very outset of the calculations\cite{SinghGelfand}.

The motivation to study such systems on frustrated lattices can be traced back to the suggestion \cite{Anderson} that the resonating-valence bond is the ground state of quantum antiferromagnets. Though there are some properties of the ground state of such systems - like the absence of long range magnetic order even down to low-temperatures - that are agreed upon, questions regarding the low-lying excitations or the gaps are not fully resolved. It is likely that such questions can be resolved through the setup of atomic quantum gases on similar optical lattices. An experimental setup for constructing and controlling optical kagome lattices using standing waves has been proposed \cite{Cirac}. Such experiments could give us a clearer understanding about the corresponding spin-models via the relevant mapping between spins and bosons\cite{Bruder}.

The Hamiltonian that describes such a model is the Bose-Hubbard Hamiltonian (BHH)\cite{ElstnerMonien},
\begin{eqnarray}
 H = -t\sum_{<i,j>}(b_i^{\dagger}b_j + \textrm{h.c}) + \frac{U}{2}\sum_{i}\hat{n_i}(\hat{n_i} - 1) - \mu\sum_{i}\hat{n_i}.\nonumber \\ \raggedleft
\end{eqnarray}
where the $b_i^{\dagger}$ and $b_i$ are bosonic creation and annihilation operators, $\hat{n_i}= b_i^{\dagger}b_i$ is the number operator, the hopping-terms $t$ are between nearest neighbors, and the system consists of a single species of soft-core bosons. The local energy term $U$ contributes to a repulsive on-site interaction between bosons and $\mu$ is the chemical potential.

Before we head onto the results, clarifying the meanings of a few technical terms is in order. A {\em topological graph} is a graph which is completely specified by its adjacency matrix, which is a representation of the interconnectedness of its vertices; a topological graph may have many distinct embeddings on the lattice (corresponding to various orientations), each of which is called a {\em cluster}. It is only required to choose one representative from these clusters (for a given topological graph) while the geometric contributions of the clusters on the lattice can be encapsulated in an appropriate lattice constant. A {\em unicyclic} ({\em bicyclic}) graph contains one (two) cycles within the graph whereas a {\em tree} is acyclic. In the rest of the paper, we call unicyclic and bicyclic graphs as {\em ring} graphs.

\section{GROUND STATE PROPERTIES}

\begin{figure}[tbp]
     \centering
    \includegraphics[width=8cm]{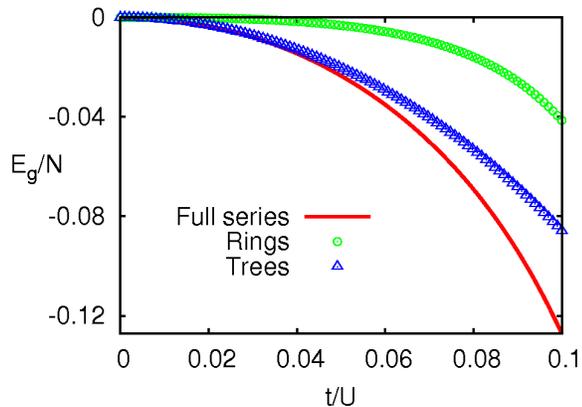}
    \caption{(Color online) Ground state energy of the Bose-Hubbard model in Equation (1) on the kagome lattice.}
    \label{GS}
\end{figure}

\begin{table}[|b|b|p|]
\begin{tabular}{lc}
\hline
Graph $c$ \T \B & Starting contribution $S(c,\mathcal{L})$\\
\hline
$P_{2}$ & $-8\lambda^2$ \\
$C_{3}$ & $24\lambda^3$ \\
$P_{3}$ & $-72\lambda^4$ \\
$T_{3,1}$ & $677.33\lambda^5$ \\
Butterfly & $-2465.78\lambda^6$ \\
$P_{4}$ & $-1777.78\lambda^6$ \\
\hline
\end{tabular}
\caption{Graphs ${c}$ with starting contribution $S(c,\mathcal{L})$ to weighted weights $L(c,\mathcal{L})W(c) > L(C_6,\mathcal{L})W(C_6)$ to the ground state energy in Equation (2).}
\label{Hex}

\end{table}
\begin{table}[|b|b|p|]
\begin{tabular}{ccc}
\hline
$t/U$ \T \B \T \B \T \B & $\alpha_{t/U}$ \T \B \T \B \T \B & $\pm\Delta\alpha_{t/U}$\\
\hline
$0.006$ & $3.729$ & $0.0017$ \\
$0.012$ & $3.033$ & $0.0049$ \\
$0.018$ & $2.623$ & $0.0095$ \\
$0.024$ & $2.32$ & $0.015$ \\
$0.03$ & $2.09$ & $0.024$ \\
$0.036$ & $1.89$ & $0.034$ \\
$0.042$ & $1.70$ & $0.045$ \\
$0.048$ & $1.53$ & $0.058$ \\
$0.054$ & $1.35$ & $0.07$ \\
$0.06$ & $1.17$ & $0.081$ \\
\hline
\end{tabular}
\caption{Exponential decay coefficients $\alpha_{t/U}$ for the correlator in Equation (3) with their corresponding errors, for varying perturbation strengths.}
\label{CorrelatorTable}
\end{table}

\begin{figure*}[tbp]
\centering
    \subfigure[ ]{
    \label{Disp}
    \includegraphics*[width=8cm]{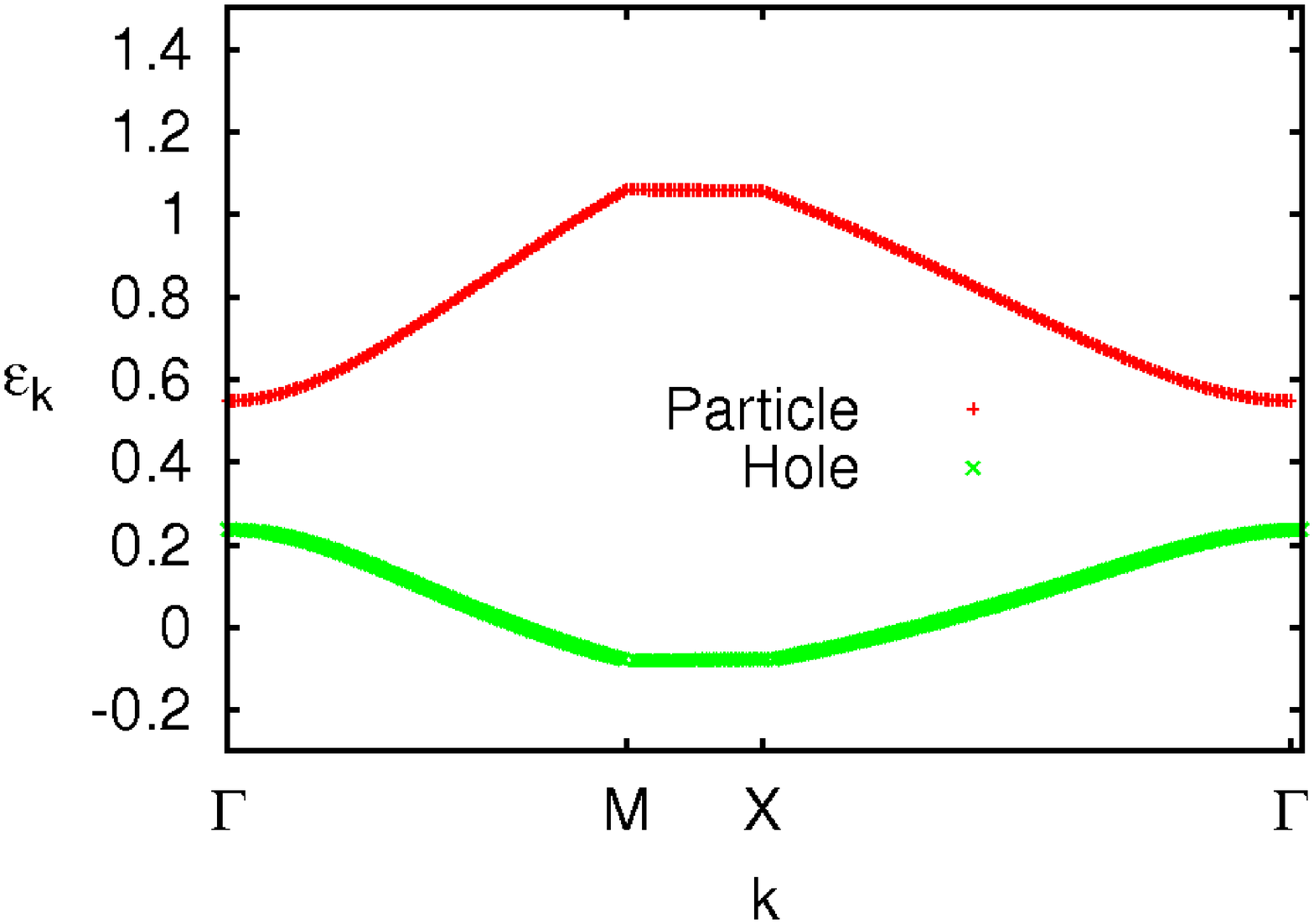}
}
    \subfigure[ ]{
    \label{Mass}
    \includegraphics[width=8cm]{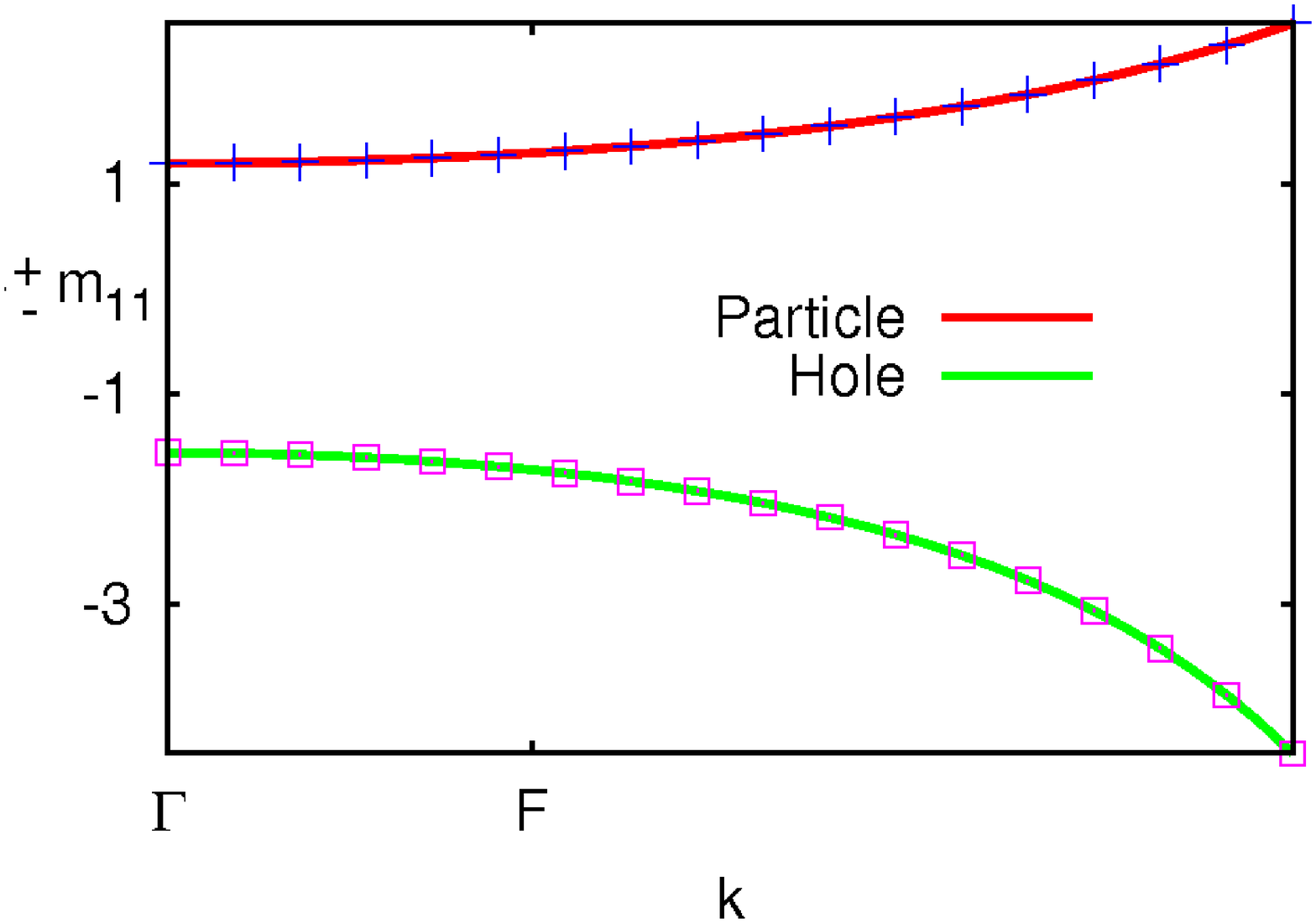}
}
    \caption{(Color online) (a) Quasiparticle dispersion for particle and hole along a given symmetry line in the first Brillouin zone for t/U = 0.050.
 (b) Effective mass tensor component $m_{11}$ for particle and hole excitations near the bottom of the band. The point $F$ is defined to be approximately one-tenth of the distance along $\Gamma - M$. The symbols indicate values obtained from numerical differentiation of a $15^\textrm{th}$ order Chebyshev-approximated dispersion $\epsilon_{\bf k}$ as in Equation (4).} \label{excited-state}
\end{figure*}

In this section, we present our results on a systematic strong coupling expansion of ground state properties. Firstly, we calculate the ground state energy per site of the system in the thermodynamic limit, to $8^{\textrm{th}}$ order in perturbation theory as
\begin{eqnarray}
\frac{E_{g}}{N}&=&-8\lambda^{2}  + 24\lambda^{3} - 40\lambda^{4} + \frac{376}{3}\lambda^{5} \nonumber\\
 & &- \frac{28640}{9}\lambda^{6} + \frac{1706768}{27}\lambda^{7}  - \frac{7644904}{9}\lambda^{8},
\end{eqnarray}
where $\lambda = -t/U$ and $N$ is the number of lattice sites. The expansion is obtained by starting the system at unit-filling and applying the hopping perturbation to the given order.

From embedding considerations on the lattice, 152 topological graphs contribute to the above expression for an $8^{\textrm{th}}$ order calculation on the kagome lattice. And from these 152 graphs, only 19 graphs have non-zero weights $W(c)$ associated with them; $W(c)$ is defined as
\begin{equation*}
 \frac{E_{g}}{N} = \sum_{c}L(c,\mathcal{L})W(c)
\end{equation*}
where the summation is over all topological graphs $c$, $L(c,\mathcal{L})$ is the lattice constant of the graph $c$ on the lattice $\mathcal{L}$ and the $W(c)$ is the weight (a corresponding polynomial) of the graph $c$. For bosonic hopping models, the lowest order of a contributing graph would be the shortest path on the graph to make a tour; this was noted previously\cite{Grzesik}, and the problem can be reformulated into the well-known Chinese Postman problem in graph theory. The validity of this optimization has been checked for ground state energy and effective Hamiltonian calculations of other lattices as well. For the same order, we compare these with the ground state calculations of the triangular and square lattices in 2D: only 38/321 topological graphs and 18/125 topological graphs contribute to these two lattices respectively, after the optimization.

We further notice that of the 19 graphs, 12 of them have closed ring structures (unicyclic and bicyclic graphs); we isolate their contributions from the trees in Fig. \ref{GS}. In the terminology of modern graph theory, the ring graphs are 3 tadpole graphs ($T_{3,1}, T_{3,2}, T_{6,1}$), 4 cycle graphs ($C_3, C_6, C_7, C_8$), a bullgraph, a butterfly graph, a 3-bellgraph, a cricket graph and graph $X_{163}$\cite{Mathworld, Rostock}. The trees are the 4 path graphs ($P_2, P_3, P_4, P_5$) and 2 star graphs ($S_3, S_4$). Thus, it is conceivable that new and interesting physics may be seen in the ground state properties of bosonic or spin models on the kagome lattice by including more than hexagonal-interaction graphs in the proposed ring-exchange model\cite{Balents, IsakovKim}. For example, the hexagon graph has, including the lattice constant, a maximal starting contribution of $-1576\lambda^6$ to the ground state energy. Graphs which have higher contributions are listed in Table \ref{Hex}.

In this ground state, we further calculate the correlator
\begin{equation}
 S_{i,j} = \langle b_j^{\dagger}b_{i} + b_{j}b_{i}^{\dagger}\rangle
\end{equation}
and determine the exponential decay coefficients i.e. $S_{i,j} \propto \mbox{exp}(-\alpha_{t/U} {\bf r}_{ij})$. We see from Table \ref{CorrelatorTable} that as we restore the $U(\textrm{1})$ symmetry and enter deeper into the MI phase, the boson particle-number spatial correlations are captured by the current method with decreasing errors.

\section{EXCITED STATE PROPERTIES}

\begin{figure}[bbp]
     \centering
    \includegraphics*[width=8cm]{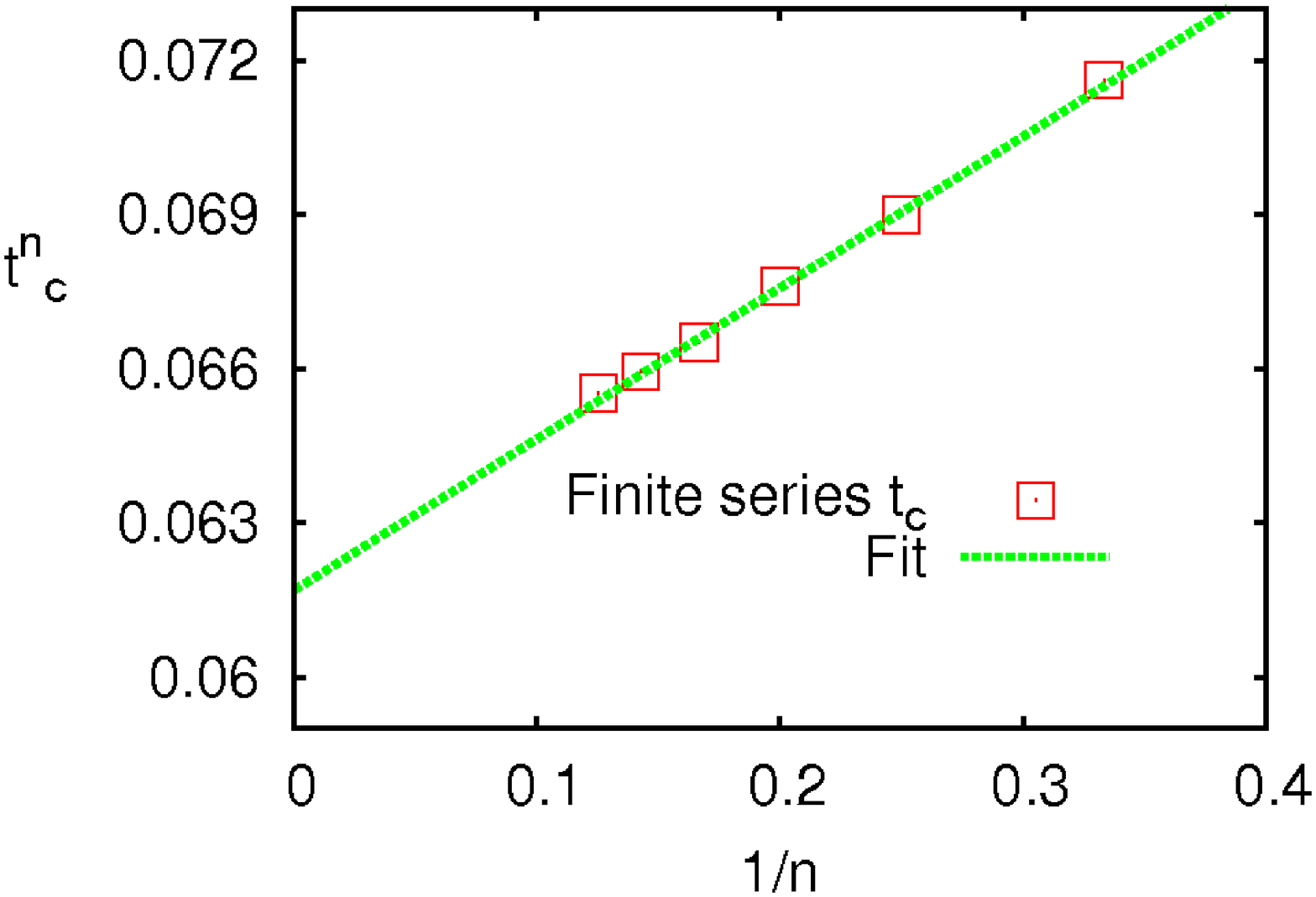}
      \caption{(Color online) (a) Extrapolation of the zeros of the truncated gap-polynomial in Equation. (5) as a function of inverse order; the errors in the plotted zeros of the finite polynomials are ignored.}
\label{Extrapolate}
\end{figure}

Single particle (hole) excitations are created within the insulating phase and their low-lying dispersion curves are obtained using the similarity transformation. The excitations on the lattice $\Delta({\bf r})$, obtained by the embedding of the effective Hamiltonians on the lattice, are diagonalized by plane-waves \cite{SinghGelfand}
\begin{equation*}
 \epsilon({\bf k}) = \sum_{{\bf r}}e^{i{\bf k}.{\bf r}}\Delta({\bf r}),
\end{equation*}
 and the spectra can be obtained for a given value of the perturbation strength. For $t/U = 0.050$, the particle and hole spectra are shown in Fig. \ref{Disp}. The smallest value of the particle-hole gap can be clearly seen at the $\bf{k} = (\textrm{0,0})$ point. We also notice a nearly flat band in the $M-X$ direction in $\bf{k}$-space.

Once we obtain the dispersion within this formalism, a straightforward calculation can give us the effective mass tensor components $m_{ij}$ for particles and holes as
\begin{equation*}
(m_{ij})^{-1} = \cfrac{\partial^{2}\epsilon({\bf k})}{\partial{k_i}\partial{k_j}}.
\end{equation*}
The mass tensor component $m_{11}$, with $k_1 \equiv k_x$, for particle and hole excitations are shown in \ref{Mass} for the same perturbation strength along a given direction in the first Brillouin zone (FBZ) near the bottom of the band. The solid lines indicate the tensor component calculated directly from the excitations $\Delta({\bf r})$. The symbols in the figure are obtained as follows: a $15^\textrm{th}$ order Chebyshev approximation of $\epsilon_{\bf k}$\cite{Press} is obtained as
\begin{equation}
 \epsilon({\bf k}) \approx \sum_{n = 0}^{N - 1} c_{n}T_{n}({\bf k}) -\frac{1}{2}c_0,
\end{equation}
where $T_{n}({\bf k})$ is a Chebyshev polynomial of degree $n$, $N = 15$ and the coefficient $c_j$ is given by
\begin{equation*}
 c_{j} = \frac{2}{N}\sum_{n = 0}^{N - 1}\epsilon \left[\cos{\frac{\pi(n + \frac{1}{2})}{N}}\right]\cos{\left (\frac{\pi j(n + \frac{1}{2})}{N} \right )}.
\end{equation*}
The approximation in Equation (4) is numerically differentiated twice to give the inverse-mass tensor components. The variables have been suitably rescaled to fit the range $[-1,1]$ where the above approximation holds. Aawy from the bottom of the band, we observe recurring divergences in the mass components within the FBZ. It is unclear how these can be renormalized into physically valid results within the current formalism.

Finally we arrive at an expression for the particle-hole gap $\delta({\bf k} = 0)$ in the insulating phase by perturbatively calculating the particle and hole contours of the ground state phase diagram and taking the difference between the two. This gives
\begin{eqnarray}
 \delta({\bf k} = 0) &=& 1 + 12\lambda - 22\lambda^2 + 75\lambda^3 - \frac{18769}{10}\lambda^4 + \nonumber \\ && \frac{1087631}{60}\lambda^5 - \frac{1133014847}{4500}\lambda^6 + \nonumber\\ &&\frac{31170197257}{13500}\lambda^7 - \frac{535354829732897}{18900000}\lambda^8. \nonumber \\ \raggedleft
\end{eqnarray}

Closing of the gap will indicate the complete disappearance of the insulating phase. To this end, two approaches are undertaken to analyze the above series. First, we truncate the above polynomial to order $n$ and solve for the zeros of the truncated polynomial and identify the relevant zero as $t_{c}^n$, with $U = 1$. This is then extrapolated to an infinite order polynomial, which should directly give an estimate of the critical point. This gives the critical point as $t_{c} = 0.0617 \pm 0.0001$ as seen in Fig. \ref{Extrapolate}.

Secondly, since the gap obeys a power law of the form\cite{ElstnerMonien}
\begin{equation*}
 \delta({\bf k} = 0) = A(t_{c} - t)^{z\nu},
\end{equation*}
where $\nu$ is the coherence-length critical exponent and $z$ is the dynamical critical exponent, we can perform a Pad\'{e} approximant to the series in Equation (5) after taking its logarithm and differentiating it, to give
\begin{equation*}
 \frac{\delta '}{\delta} = \frac{A'(t)}{A(t)} + \frac{z\nu}{t_{c} - t}.
\end{equation*}

Considering the diagonal approximants ([2/2], [3/3], [4/4]) we find $t_c = 0.0615 \pm 0.0001$ and $\nu = 0.651 \pm 0.006$ after setting $z = 1$ at the multicritical point\cite{Fisher}. The critical point calculated by the direct extrapolation is seen to be a reasonably good estimate for an $8^{\textrm{th}}$ order calculation; the critical exponent is to be compared to the corresponding value of the 3D XY model, which has $\nu = 0.67155 \pm 0.00027$\cite{Campostrini}. It is interesting to note that for unit-filling on the kagome lattice, performing a mean-field calculation\cite{Sachdev} gives $t_c \approx 0.04289$. For comparison, we compute the superfluid stiffness $\rho$ using the ALPS software\cite{ALPS1, ALPS2} to perform a Quantum Monte Carlo simulation implemented by the worm algorithm for 3 different lattice sizes (number of kagome unit cells: $3$x$3$, $6$x$6$, $9$x$9$), with a maximum of 2 bosons per site. Using the scaling {\em ansatz}\cite{Cha}
\begin{equation*}
\rho = L^{2-d-z}\tilde{\rho}(aL^{\frac{1}{\nu}}(t_{c}/t-1)), 
\end{equation*}
we get an estimate  $t_{c} \approx 0.061$. Here, $d$ is the dimensionality, $L$ is the linear system size, $\tilde{\rho}$ is the scaling function and $a$ is a non-universal metric factor. As before, we set $z=1$ at the multicritical point.

To conclude, we have seen that the number of contributing graphs to ground state energy calculations on the kagome lattice using the series expansion is only a small fraction of the number calculated from geometric considerations of embedding, which is a feature of hopping models. The optimization for such models, resulting from the use of the Chinese Postman algorithm to eliminate graphs, has been checked. Furthermore, it is possible that the inclusion of these ring graphs into the interaction terms of the previously proposed ring-exchange Hamiltonian might give further insight into the ground state properties of similar models defined on the kagome lattice. We also present an accurate determination of the critical point and critical exponent for the model along with other excited state properties, albeit the perturbative order was lower than previously studied 2D lattices; yet we believe that this is a reasonably precise estimate that can be used as a reference for future work.

\acknowledgments

We thank Rajiv R. P. Singh for helpful discussions especially regarding the question of graph contribution for hopping models, which we later found to be an unpublished but known fact. Reference data for correlators from Niklas Teichmann, and discussions with Simon Trebst and Matthias Troyer are gratefully acknowledged. VKV is partially supported by the Bonn-Cologne Graduate School within the DFG's German Excellence Initiative.

\end{document}